\documentstyle{aipproc}
                                       
\begin{document}

\title{Pre-Big-Bang in String Cosmology}

\author{M\'onica Borunda and M. Ruiz Altaba}
\address{Departamento de F\'{\i}sica Te\'orica\\Instituto de F\'{\i}sica\\
Universidad Nacional Aut\'onoma de M\'exico\\
Apartado Postal 20-364\\01000 M\'exico, D.F. }

\maketitle

\begin{abstract}

We compute the amount of inflation required to solve the horizon problem of cosmology in the pre-big-bang scenario. First  we give a quick overview of string cosmology as  developed by Veneziano and collaborators. Then we show that the amount of inflation
  in this background solves the horizon problem. We discuss fine-tuning.
\end{abstract}

%
%
%
%

The standard  cosmological model works well at late times, explaining the red shift, the cosmic microwave background and the cosmic primordial
nucleosynthesis, but it has  problems associated with the initial
singularity, the homogeneity, the isotropy, the flatness,  and the large-scale
structure.
Inflation   solve these  problems except for the initial singularity. Inflationary models are  constrained by demanding a graceful exit, the  right  amount of reheating and the right amount of large-scale inhomogeneities. This requires fine-tuned initial 
conditions and inflation potentials. But inflation does not even attempt to solve the initial singularity problem.

A few years ago a stringy cosmology was built with the very basic postulate that the universe did indeed start near its trivial vacuum, solving the initial singularity problem\cite {uno}.

String theory is the only consistent theory containing  quantum gravity. Each of the normal modes of vibration of a quantum string  is  a conventional particle. At large distances ($\lambda >> 10^{-33}$cm), strings appear as particles.  In string theory t
here are symmetries known as dualities \cite {uno}\cite{dos}\cite{tres}\cite{cuatro} which allow us to find two equivalent solutions to the problem of the initial singularity.

The low-energy effective action for bosonic closed strings is:

\begin{equation}
S={1\over 4\pi\alpha '}\int d^4x \sqrt{{\rm det}g\,}\, e^{-\phi} ( R + \partial_\mu\phi\partial^\mu\phi + ... )
\end{equation}
 Where $R$ is the Ricci scalar,  $(\alpha ')^{-1}$ is the string tension, $g_{\mu\nu}$ is the metric and $\phi$ is  the so-called dilaton, a scalar massless particle which may play the  role of a inflaton.

The  equations of motion for this action in the flat F.R.W. metric

\begin{equation}
ds^2=dt^2-a^2(t)[dr^2+r^2d\theta^2+r^2\sin^2\theta d\varphi]
\end {equation}
are invariant under a stringy symmetry known as ``scale factor duality":

\begin{equation}
t\Rightarrow -t\qquad a(t)\Rightarrow \tilde{a}(-t)=a^{-1}(-t)\qquad \phi\Rightarrow \phi -\ln \sqrt{|g|}
\end{equation}
We simulate matter with a  perfect fluid stress-energy tensor $T^\nu_\mu ={\rm diag}(\rho\,  ,-p\delta^j_i)$ where $\rho$ is the energy density and $p$ is the pressure. Furthermore, we set $\phi={\rm cte}=\phi_0$ and  $V(\phi )=0$. For  radiation  we get 
for $t>0$ our universe in its radiation dominated epoch:

\begin{equation}
a=\left ({t\over t_0} \right )^{1\over 2}\qquad \phi=\phi_0
\end {equation}
\begin{equation}
\rho =3p=\rho_0\left ({t\over t_0} \right )^{-2}\qquad G\sim\alpha 'e^{\phi_0}={\rm constant}
\end{equation}
Applying the duality transformation, we find for $t<0$:

\begin{equation}
a=\left ( -{t\over t_0} \right )^{-1/2}\qquad \phi =\phi_0 -3\ln \left ( -{t\over t_0} \right )
\end{equation}
\begin{equation}
\rho =-3p=\rho_0\left( -{t\over t_0} \right )\qquad G\sim \alpha ' e^{\phi_0}\left (-{t\over t_0} \right )^{-3}\neq {\rm constant}
\end{equation}

The universe starts at $t=-\infty$ with flat empty Minkowski space with zero coupling; Newton's coupling then starts growing  and the universe inflates  non-exponentially  $t<<0$.  For $t>>0$ we recover flat Minkowski space with a weak coupling and a dece
lerated expansion (which corresponds to our universe). This is a description of the evolution of our universe at times well before and after the big bang ($t=0$), but what happens during the high curvature regime is, of course, unknown.  We may try to sim
ulate it with a time dependent equation of state such as

\begin{equation}
p=\gamma (t)\rho
\end{equation}
where $-1/3<\gamma (t) <1/3$.There is an expanding solution \cite{dos} ($H>0$) for all $t$: the universe dominated by string  matter ($p=-\rho /3$), starts from flat space ($H\rightarrow 0$), an unstable solution, with  weak coupling ($e^\phi\rightarrow 0
$) regime evolving through an inflationary phase $[a(t)\sim (-t)^{-1/2}]$  phase with  gravitational coupling ($e^\phi ={\rm const.}$).
An analogous solution exists, in which the universe is always  contracting ($H<0$) in correspondence with the dual equation of state $\gamma(-t)$.

What about the other dimensions that string theory allows?

Consider a background in which, during the pre-big-bang phase ($t<0$),  $d$ dimensions expand with scale factor $a(t)$ while $n$ dimensions shrink with scale factor $a^{-1}(t)$ with an equation of state $p=-q=-\rho /(d+n)$.

\begin{equation}
g_{\mu\nu}={\rm diag }(1,-a^2(t)\delta_{ij},-a^{-1}(t)\delta_{ab})
\end{equation}
The solution is $a(t)\sim (-t)^{-2/(d+n+1)}$, $t<0$\cite {dos}. This background evolves into a phase of maximal, finite curvature, after which it approaches the dual, decelerated regime ($t>0$) in which the internal dimensions are not frozen, but keep con
tracting like $a^{-1}(t)\sim t^{-2/(d+n+1)}$ for $t\rightarrow +\infty$. The dilaton vacuum expectation value  does not settle down to a finite constant value after the big-bang, but tends to decrease during the phase of decreasing curvature. Such a decre
ase of $\phi$ is driven by the decelerated shrinking of the internal dimensions which are not frozen.

\section*{Constraints on initial conditions}

The condition to solve the horizon problem (in the Einstein frame, thus the tildes) is 

\begin{equation}
d_{{\rm HOR}}(t_f)={\tilde a}(t_f)\int_{t_i}^{t_f}{dt'/{\tilde a}(t')}>{\tilde a}(t_f)H_0^{-1}/{\tilde a}_0
\end{equation}
 where $H_0^{-1}$ is the size of the observed Universe ($H_0^{-1}\sim 10^{28}$cm),    $t_f$ is the time by which the horizon problem is solved and $t_i$ the time when inflation begins. For us, $t_i$ and $t_f$ determines
the time range when the pre-big bang description  remains valid.
Obviously, letting $t_i\to - \infty$ we see   that the horizon problem is solved both for $k=0$ and for $k=-1$.   Still, it is of some 
interest to ask   how long did the universe have to behave stringily before the
big bang in order for it to come out free of flatness and horizon problems from
the high curvature epoch (around $t=0$) \cite{seis} \cite{siete}.

The amount of expansion required to solve the horizon problem is given  by the ratio

\begin{equation}
Z={H(t_f)a(t_f)\over H(t_i)a(t_i)}
\end{equation}
Experimentally (or rather, observationally), we need

\begin{equation}
Z>e^{60}
\end{equation}
in order to solve the horizon problem for our big universe.

Since our  effective actions stops being valid when  gravity becomes   strongly coupled, we expect the pre big bang inflationary epoch to be over by  the time $t_f$ when 

\begin{equation}
e^{-\phi (t_f)}>> 1
\end{equation}
Similarly, the same effective actions remain valid only  while the curvature is not too big:

\begin{equation}
 H^{-1}(t_f )\sim (-t_f)>>l_{st}
\end{equation}
When $k=0$,  the amount of inflation is thus

\begin{equation}
Z=\left ({-t_i\over l_{st}} \right )^{3/2}
\end{equation}
This give us that $t_i<-10^{17}l_{st}$ in order to get the  amount of inflation for  succesfully solve the horizon problem.

We conclude from string cosmology that: (a) inflation comes naturally, without ad-hoc fields, (b) initial conditions are natural, (c) the kinematical problems of the standard cosmological model are solved and  (d) a hot big bang could be a natural outcome
 of our inflationary scenario. Furthermore, it can be shown that (e) perturbations do not grow too fast to spoil homogeneity, (f) our understanding of the high curvature (stringy) phase is still poor and (g) the amount of inflation required needs some fin
e-tuning of initial conditions for $k=0$ and  for $k=-1$ in order to solve the horizon problem.

This work is supported in part by the project DGAPA IN103997.

\end{document}